\documentclass[twocolumn,letterstyle]{aa}
\usepackage{graphicx}
\usepackage{txfonts}
\begin{document}
   \title{Extreme helium stars: non-LTE matters} 
   \subtitle{Helium and hydrogen spectra of the unique objects V652\,Her and HD\,144941}


   \author{N. Przybilla\inst{1}
          \and
          K. Butler\inst{2}
	  \and
	  U. Heber\inst{1}
	  \and
	  C.S. Jeffery\inst{3}
          }

   \offprints{N. Przybilla (przybilla@sternwarte.uni-erlangen.de)}

   \institute{Dr. Remeis-Sternwarte Bamberg, 
Sternwartstrasse 7, D-96049 Bamberg,
Germany
\and
Universit\"atssternwarte M\"unchen, Scheinerstrasse 1, D-81679 M\"unchen,
Germany
\and
Armagh Observatory, College Hill, Armagh BT61 9DG, Northern Ireland
}

   \date{Received; accepted}

   \abstract{Quantitative analyses of low-mass hydrogen-deficient
(super-)giant stars -- so-called extreme helium stars -- to date
face two major difficulties. First, theory fails to reproduce the observed
helium lines in their entirety, wings {\em and} line cores. Second, a general 
mismatch exists for effective temperatures derived from ionization equilibria and from
spectral energy distributions. Here, we demonstrate how the issue can be
resolved using state-of-the-art non-LTE line-formation
for these chemically peculiar objects. 
Two unique high-gravity B-type objects are discussed in detail, the pulsating variable 
\object{V652\,Her} and the metal-poor star \object{HD\,144941}. In the first case 
atmospheric parameters from published LTE analyses are largely recovered, in the
other a systematic offset is found. Hydrogen abundances are
systematically smaller than previously reported, by up to a factor $\sim$2.
Extreme helium stars turn out to be important testbeds for non-LTE model atoms for helium. 
Improved non-LTE computations show that analyses assuming LTE or
based on older non-LTE model atoms can predict equivalent widths, 
for the \ion{He}{i}\,10\,830\,{\AA} transition in particular, in error by up to a factor 
$\sim$3. 
   }

   \maketitle


\section{Introduction}
Extreme helium stars (EHes) are a rare class of hydrogen-deficient objects 
showing spectral characteristics of early A to late O-type (super-)giants. 
Their chemical composition is domi\-nated by fusion products from the CNO-cycle 
and/or the 3$\alpha$-process. Therefore, EHes provide important clues
for the study of nuclear astrophysics. 
Most of the two dozen known EHes could be explained by post-AGB
evolution, linking R\,Cr\,B stars to Wolf-Rayet type central stars of planetary
nebulae, see Heber~(\cite{Heber86}) and Jeffery~(\cite{Jeffery96}) for reviews. 

V652\,Her and HD\,144941 are unique among the class members in several aspects.
The carbon, nitrogen and oxygen abundances of both stars are consistent with being processed
through the CNO cycle, while all other EHes display C-rich material at the
surface, indicating that 3$\alpha$-processing has occurred in addition.    
Both stars have gravities too large for post-AGB evolution. 
Saio \& Jeffery~(\cite{SaJe00}) suggest that V652\,Her may be the result of
a He$+$He white dwarf binary merger.
HD\,144941 
is also outstanding because of its strong metal-deficiency, larger than observed in any other 
EHe star. 


Because of their peculiar chemical composition 
EHes are important testbeds for stellar atmosphere modelling. 
Since helium is by far the most abundant element, the He line spectrum
can be studied in more detail than in any other type of star, 
e.g. in HD\,144941 all forbidden transitions of \ion{He}{i} 
can be measured. 
Hydrogen is deficient by a factor of 100 or more and therefore the Balmer
lines are very weak, unlike in any other B-type~star.


Several LTE spectral analyses encountered two difficulties 
to be most troublesome: {\sc i)} synthetic spectra have so far not 
succeeded in matching the observed 
helium lines in their entirety, and {\sc ii)} spectroscopic and spectrophotometric
temperatures differ systematically.
As inadequacies in the basic parameter determination can potentially 
hamper any further interpretation, the issue needs to be 
resolved.
In the following we investigate the steps that need to be taken to improve the
modelling, and the consequences of these for our understanding of EHe stars. 


\section{Model calculations}
  \begin{figure*}
   \centering
\includegraphics[width=.842\textwidth]{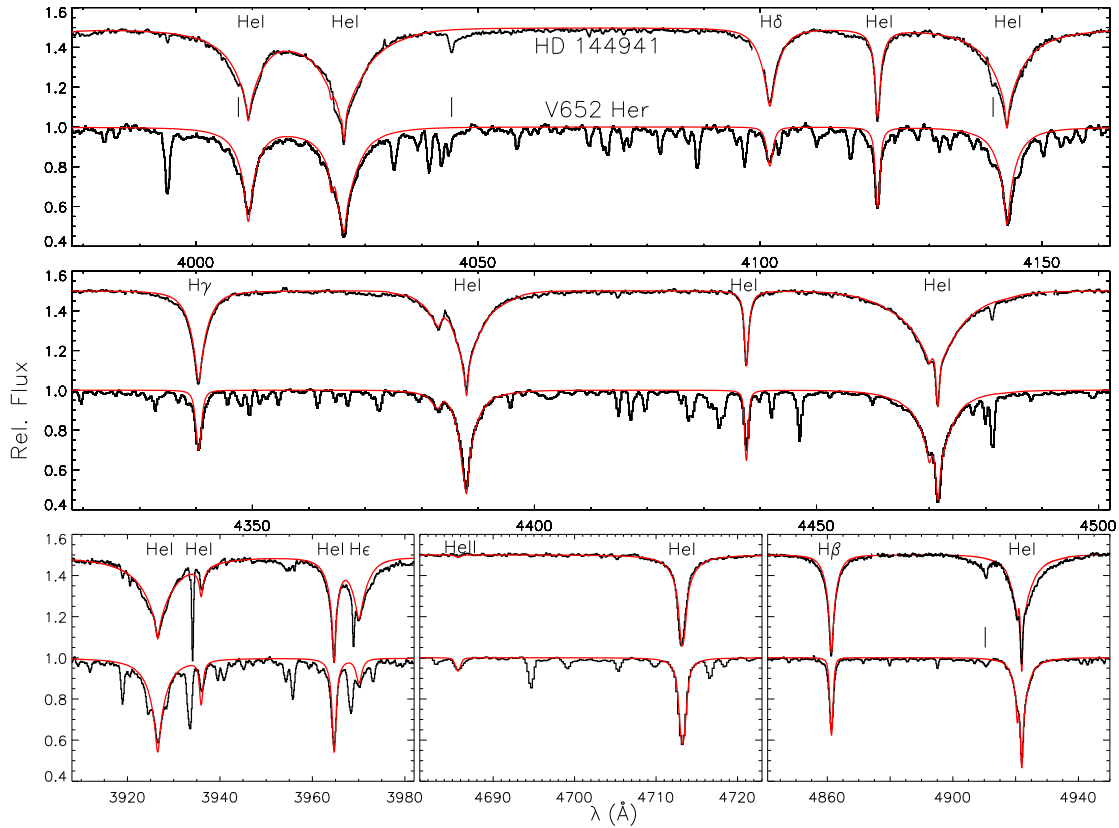}
\caption{Comparison of the normalised spectra of V652\,Her (lower) 
and the metal-poor hydrogen-deficient star 
HD\,144941 (upper histogram) with our non-LTE spectrum synthesis for
helium and hydrogen (full red line): excellent agreement for the entire line
profiles -- wings {\em and} line cores -- is found. The He and H features have
been labelled. Forbidden components of \ion{He}{i} lines missing in our
modelling are indicated by short vertical marks. The red wing of \ion{He}{i}
4922\,{\AA} in HD\,144941 may be affected by an artifact in the data
reduction.
              }
         \label{vis}
   \end{figure*}

Our model calculations are carried out in a hybrid approach, as commonly
used for analyses of analogous B-type stars with normal chemical composition. 
First, hydrostatic,
plane-parallel and line-blanketed (via an opacity sampling technique) LTE
model atmospheres are computed with the {\sc Atlas12} code (Kurucz~\cite{Kurucz96}).
Note that we have replaced the photoionization data for \ion{He}{i} levels
with principal quantum number $n$\,$=$\,2 as used by Kurucz with data from
the Opacity Project (Fernley et al.~\cite{Fernleyetal87}). In particular the
cross-sections for the $2p$\,$^3P^{\circ}$ level are increased by a
factor $\sim$2 at threshold, thus improving the fits of computed energy
distributions with observation. Stellar parameters and elemental abundances 
from Jeffery et al.~(\cite{Jefferyetal01}, JWP01) and Harrison
\& Jeffery~(\cite{HaJe97}, HJ97) are adopted for the initial models of 
the stars. 
Details on the observations can also be found there,
which are complemented by further data from Jeffery et
al.~(\cite{Jefferyetal99}).

Then, non-LTE line formation is performed on the resulting model stratifications. 
The coupled radiative transfer~and~statis\-tical equilibrium equations are
solved and spectrum synthesis with refined line-broadening theories is 
performed~using~{\sc Detail} and {\sc Surface} (Giddings~\cite{Giddings81}; 
Butler~\&~Giddings~\cite{BuGi85}).~Both codes have undergone major revisions and
improvements~over the past few years. State-of-the-art model atoms for 
hydrogen (Przybilla \& Butler~\cite{PrBu04}, PB04) and helium
(Przybilla~\cite{Przybilla05}, P05) are utilised, which differ from older
model atoms mostly by use of improved collision data from {\em ab-initio}
computations for electron impact excitation processes. For \ion{He}{i}
these are the data from Bray et al.~(\cite{Brayetal00}) and Sawey \&
Berrington~(\cite{SaBe93}), and for \ion{H}{i} the data from computations by
Butler, as summarised in PB04 -- see this work and P05 for further details on the
modelling. For comparison, additional calculations are made
using the helium model atom of Husfeld et al. (\cite{Husfeldetal89},\,HBHD89).
The theory of Barnard et al.~(\cite{Barnardetal69}) 
which is discussed further by Auer \& Mihalas~(\cite{AuMi72}) 
is utilised for a realistic description of line
broadening, supplemented by data from 
Dimitrijevi\'c \& Sahal-Br\'echot~(\cite{DiSa90}).

\begin{table}
\caption{Photospheric parameters of the sample stars}
\label{table1}
\vspace{-2mm}
\centering
\begin{tabular}{lr@{$\pm$}lr@{$\pm$}l}
\hline\hline
 & \multicolumn{2}{c}{V652\,Her\,($R_{\rm max}$)} & \multicolumn{2}{c}{HD\,144941}\\
\hline
$T_{\rm eff}$\,(K)                  & 22\,000 & 500               & 22\,000 & 1\,000\\
$\log g$\,(cgs)                     &    3.20 & 0.10              & 4.15    & 0.10\\
$\xi$\,(km\,s$^{-1}$)               &       4 & 1                 & 8       & 2\\
$n_{\rm H}^{\rm NLTE}$\,(by number) &   0.005 & 0.0005            & 0.035   & 0.005\\
\hline
\end{tabular}
\end{table}

The spectrum synthesis is compared to observation in order to derive
improved values for the stellar parameters and hydrogen and helium abundances, 
giving `best fits' in an iterative approach. Effective temperatures $T_{\rm eff}$ 
are determined from the \ion{He}{i/ii} non-LTE ionization equilibrium 
and the Stark-broadened \ion{He}{i} lines act as surface gravity indicators.
Stellar parameters of the final models (with estimated uncertainties) 
are summarised in Table~\ref{table1}, including microturbulence $\xi$.
For V652\,Her the atmospheric
parameters agree very well with those found by JWP01, 
except for the hydrogen abundance $n_{\rm H}$, which is reduced by a factor
$\sim$2 (see Sect.~\ref{hydrogen}). For HD\,144941, however, the resulting
atmospheric 
parameters differ significantly from previous work of HJ97,
implying a reduction in $T_{\rm eff}$ by 1\,200\,K and an increase in
surface gravity by a factor $\sim$2. The reduction in the hydrogen
abundance is less pronounced~in~this~star.


\section{Testbed for non-LTE spectrum synthesis}


\subsection{Optical helium line spectrum}

As can be seen from Fig.~\ref{vis} excellent agreement of the non-LTE 
spectrum synthesis with the observed line
profiles of the He features -- wings {\em and} cores (and forbidden
components)\,--\,is finally obtained in our approach. 
Similar agreement of theory with observation in the {\em visual} spectral
range is obtained using the HBHD89 model atom. The great
improvement achieved becomes obvious when comparing Fig.\,\ref{vis}~to~Fig.\,2
of HJ97 and Fig.\,5 of JWP01. 
This resolves one of the most persistent inconsistencies in quantitative 
analyses of EHes. 
Only a few predicted forbidden components (Beauchamp \& Wese\-mael~\cite{BeWe98})
are missing because the broadening data are unavailable to us. 

   \begin{figure}
   \centering
   \includegraphics[width=.47\textwidth]{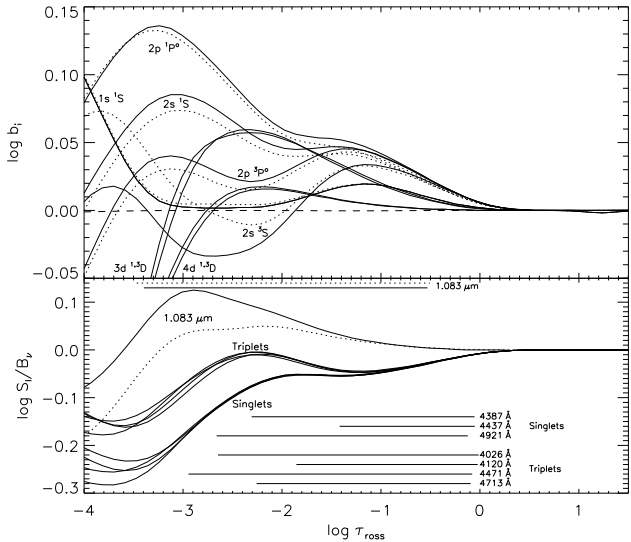}\\[-4mm]
      \caption{Departure coefficients $b_i$ for \ion{He}{i} in 
the model of V652\,Her as a function of
Rosseland optical depth $\tau_{\rm ross}$ for the $n$\,$=$\,1 and 2 and
selected $n$\,$=$\,3 and 4 levels from computations 
using the P05 and the HBHD89 model atoms (upper panel, full and dotted 
lines, respectively).
The \ion{He}{ii} ground state is also indicated (dashed line). In the lower
panel ratios of line source function $S_l$ to Planck function
$B_{\nu}$ for several \ion{He}{i} lines are displayed. Line
formation depths are indicated. The considerable differences in the 
populations of the $2s$~$^3S$ state in the two model atoms translate to a
discrepant $S_l/B_{\nu}$ ratio for the \ion{He}{i} 1.083\,$\mu$m
transition, giving a substantially weakened line for the modern model atom, 
while computations based on the old model deviate only little
from detailed balance. 
              }
         \label{dep}
   \end{figure}

The \ion{He}{i} lines in the visual experience non-LTE
strengthening, facilitated by the non-LTE overpopulation of the $n$\,$=$\,2
states relative to the levels of higher principal quantum number, see
Fig.~\ref{dep}. There, the runs of non-LTE departure coefficients
$b_i$\,$=$\,$n_i/n_i^*$ (the $n_i$ and
$n_i^*$ being the non-LTE and LTE populations of level $i$, respectively)
and line source functions $S_l$ (relative to the Planck function $B_{\nu}$)
are displayed. The non-LTE overpopulation occurs because of recombinations to levels
of \ion{He}{i} at higher excitation energies and subsequent de-excitation via
downward cascades to the (pseudo-)metastable 
$2s$ states (the singlet resonance lines are close to detailed balance). 
The singlet lines are in general subject to larger non-LTE effects than the
triplet lines. 
Note that the level populations deviate by only a few percent from detailed equilibrium
at the formation depths of the continua, showing that the assumption of
LTE for the model atmosphere computations is appropriate.

\subsection{Optical hydrogen line spectrum}\label{hydrogen}
\begin{figure}
   \centering
   \includegraphics[width=.47\textwidth]{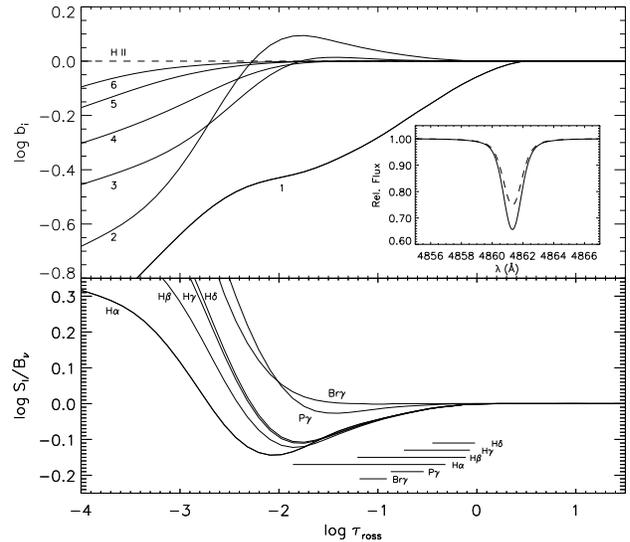}\\[-4mm]
      \caption{Same as Fig.~\ref{dep}, but for \ion{H}{i}. The $b_i$ 
are labelled by the principal
quantum number. Non-LTE (full line) and LTE profiles (dashed line) of
H$\beta$ are compared in the inset. Line source-functions
are displayed for Balmer lines and Paschen and Brackett
lines in the $J$- and $K$-band.
              }
         \label{deph}
   \end{figure}

In analogy to helium hydrogen can also be supposed to experience non-LTE effects.
Departure coefficients for several energy levels and 
line source-functions for selected \ion{H}{i} lines are displayed in
Fig.~\ref{deph}. 
The \ion{H}{i} ground state is strongly depopulated, however at
depths irrelevant for the formation of the Lyman continuum. Consequently,
non-LTE effects on H are also unimportant for the atmospheric
structure calculations.
A small overpopulation of the first excited
level occurs at line-formation depths. All other levels of higher $n$ 
are closely coupled to the continuum (which is in detailed balance) at 
line-formation depths. This results in a non-LTE strengthening of the Balmer
lines, with non-LTE and LTE equivalent widths for e.g. H$\beta$ differing by 
more than 30\%. A reduction of H abundances 
by a factor up to $\sim$2 is indicated relative to previous LTE studies 
(for V652\,Her the lines are formed on the flat part of the curve of growth).  

\subsection{Infrared line spectra}\label{infrared}


Spectral lines in the Rayleigh-Jeans tail of the spectral 
energy distribution, e.g. in the near-IR for EHes, can experience 
amplified non-LTE effects (see e.g. PB04). Analyses in the near-IR range 
are highly useful for constraining the atomic data input for the non-LTE
computations, \ion{He}{i} in the present~case.

Non-LTE weakening is indicated from Fig.~\ref{dep} for the $2s$~$^3S$--$2p$~$^3P^{\rm o}$
transition in the $J$-band, and an emission feature in the case of the
$2s$~$^1S$--$2p$~$^1P^{\rm o}$ transition in the $K$-band. Here,~the upper 
levels are overpopulated relative to the lower states. The \ion{He}{i}
1.083\,$\mu$m transition turns out to be highly sensitive to the
details of the model calculations, in particular to the use of accurate
photoionization cross-sections for the $2s$~$^3S$ level, and the account of
line blocking. As is the case of early-type main sequence stars (P05),
large differences to computations based on older \ion{He}{i} model atoms such as
those of HBHD89 are found. The state-of-the-art model atom predicts the equivalent
width $W_{\lambda}$ of \ion{He}{i} $\lambda$1.083\,$\mu$m to be a factor 
$\sim$3 smaller than the old model atom, which indicates only small deviations 
from detailed balance.

\begin{figure}
   \centering
   \includegraphics[width=.47\textwidth]{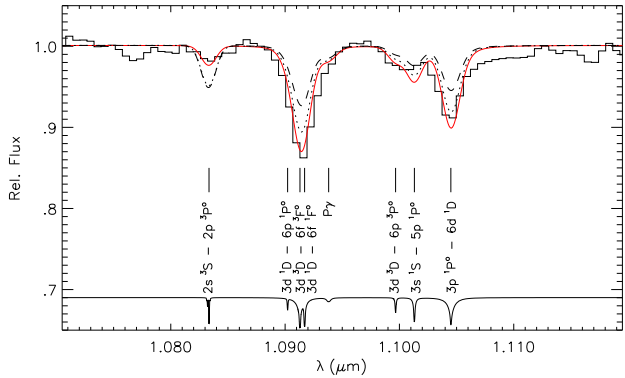}\\[-2mm]
      \caption{Comparison of the $J$-band spectrum 
of V652\,Her (histogram) with 
three model predictions: LTE (dashed line) and two non-LTE computations using the
model atom of HBHD89~(dotted line) and the improved model atom
of P05~(full red line). In contrast to the other
transitions, \ion{He}{i} $\lambda$1.083\,$\mu$m experiences 
strong non-LTE weakening. An unbroadened synthetic spectrum is displayed
at the bottom 
in order to illustrate what can be expected from high-resolution observation.
              }
         \label{ir}
\end{figure}

The significance of the predictions based on the new model atom is
demonstrated in Fig.~\ref{ir}, where a comparison is made with the only
published spectrum for an EHe star in the near-IR.
Good agreement of the model computations with the intermediate-resolution
$J$-band observation (JWP01) is found,
containing in addition many strong \ion{He}{i} lines originating from
$n$\,$=$\,3 states. Note however,
that the line broadening is dominated by instrumental effects.
High-resolution observations would be highly desirable in order to 
constrain the modelling even further. The analogous 
$2s$~$^1S$--$2p$~$^1P^{\rm o}$ transition in 
the $K$-band at 2.059\,$\mu$m is predicted to be entirely in emission, with
$W_{\lambda}$\,$\simeq -$\,200\,m{\AA} 
for V652\,Her. The non-LTE effects are smaller for HD\,144941 because of 
higher atmospheric densities, implying a lower ionization of the 
atmospheric helium.
Finally, we find that the hydrogen Paschen and Brackett series are much
closer to detailed equilibrium than the Balmer lines.

\section{Spectral energy distribution}

\begin{figure}
   \centering
   \includegraphics[width=.48\textwidth]{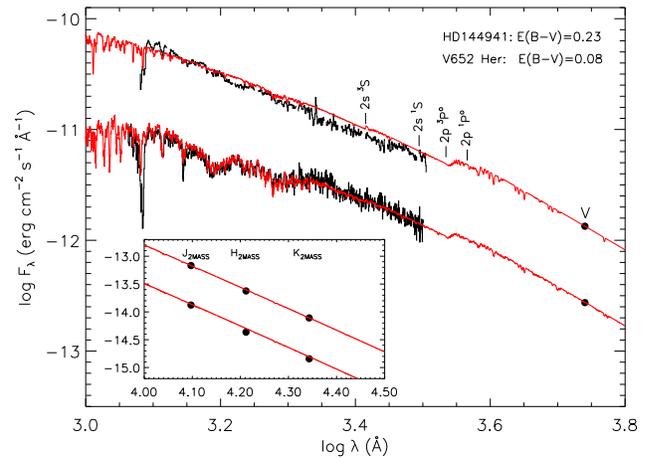}\\[-2mm]
      \caption{Fits of theoretical energy distributions (full red
lines) to IUE, visual and IR spectrophotometry for V652\,Her 
(near maximum radius, lower jagged
curve and dots) and HD\,144941 (upper curve). The observed
fluxes are dereddened~by~the amounts indicated and the HD\,144941 data
are shifted vertically by $+$0.3 units. The models are normalised to the observed 
John\-son $V$ magnitudes. Locations for the bound-free edges of the
$n$\,$=$\,2 levels of \ion{He}{i} are indicated.
              }
         \label{flux}
   \end{figure}

The second major difficulty in the modelling of EHes~is~a~gen\-eral mismatch
of effective temperatures derived from ionization equilibria and
spectrophotometry.
We~compare~the~{\sc Atlas12} model fluxes for stellar parameters 
derived from the~\ion{He}{i/ii}~ion\-ization equilibria (see Table\,\ref{table1}) 
with IUE spectrophotometry and visual and near-IR photometry in Fig.~\ref{flux}. 
Excellent agreement is found for V652\,Her, and
reasonable agreement~for HD\,144941 when interstellar reddening is accounted
for. 

The fit for HD\,144941 may be improved by allowing for
changes of the stellar metallicity and/or the effective temperature. 
Also, a non-standard reddening law cannot be ruled out. A completely
satisfactory solution for this metal-poor object is beyond the scope of the 
present paper,
as metal abundances have to be determined, and $T_{\rm eff}$ should be
verified from non-LTE ionization equilibria of metals. However, model atoms
for several of the strategic species are unavailable at present. 

\section{Conclusions}
The present work demonstrates how a hybrid non-LTE approach 
using line-blanketed LTE model atmospheres 
and non-LTE spectrum synthesis succeeds in solving the most 
persistent problems in the quantitative spectroscopy of EHes. Test
computations for two objects imply that stellar parameters from classical
analyses of EHes may be subject to systematic modification in the hybrid
non-LTE approach. Lower hydrogen abundances than reported previously are
found, by up to~a~factor~$\sim$2.

The atmospheres of the chemically peculiar EHes allow non-LTE
model atoms in an environment complementary to normal stars to be tested.
The present work verifies the inadequacy of
LTE and older non-LTE model atoms for helium. Only a recently improved 
model atom (P05) reproduces
observation for all available spectroscopic indicators, including the
\ion{He}{i} $\lambda$1.083\,$\mu$m feature in particular, which changes by a
factor $\sim$3 in equivalent width while the lines in the
visual remain practically unaffected.
High-resolution spectroscopy in the near-UV and near-IR will
be required to put even tighter constraints on the atomic model for one of the 
most 
basic elements. Finally, an improved understanding of EHes in the context of 
nuclear
astrophysics will require non-LTE studies of metal abundances.




\end{document}